\documentclass[twocolumn]{article}
\usepackage[pdftex]{graphicx}
\usepackage{ragged2e}
\usepackage{array}
\newcolumntype{P}[1]{>{\centering\arraybackslash}p{#1}}
\newcolumntype{M}[1]{>{\centering\arraybackslash}m{#1}}
\usepackage{multirow}
\usepackage{cite}
\usepackage{xspace}
\usepackage[]{subfig}

\usepackage{newtxtext}
\usepackage[varg]{newtxmath}
\usepackage{algorithm,algpseudocode}
\usepackage{caption}
\DeclareGraphicsExtensions{.pdf,.png,.jpg,.jpeg}

\usepackage{xspace}
\newcommand{\itutp}{P.1203\xspace}
\newcommand{\itut}{ITU\nobreakdash-T}

\usepackage[usestackEOL]{stackengine}

\usepackage{siunitx} 
\makeatletter
\providecommand\add@text{}
\newcommand\tagaddtext[1]{%
	\gdef\add@text{#1\gdef\add@text{}}}%
\renewcommand\tagform@[1]{%
	\maketag@@@{\llap{\add@text\quad}(\ignorespaces#1\unskip\@@italiccorr)}%
}
\makeatother

\usepackage{authblk}
\usepackage[colorlinks=true, urlcolor=black]{hyperref}
\begin{document}
	\title{A Study on Impacts of Multiple Factors on\\ Video Qualify of Experience}
	\author[1]{Huyen~T.~T.~Tran}
	\author[2]{Nam~Pham~Ngoc}
	\author[1]{Truong~Cong~Thang}
	\affil[1]{The University of Aizu, Aizuwakamatsu, Japan}
	\affil[2]{VinUniversity, Vietnam}
	\date{}                     
	\setcounter{Maxaffil}{0}
	\renewcommand\Affilfont{\itshape\small}

	\maketitle

	\begin{abstract}
		HTTP Adaptive Streaming (HAS) has become a cost effective means for multimedia delivery nowadays. However, how the quality of experience (QoE) is jointly affected by 1) varying perceptual quality and 2) interruptions is not well-understood yet. In this paper, we present the first attempt to quantitatively quantify the relative impacts of these factors on the QoE of streaming sessions. To achieve this purpose, we first model the impacts of the factors using histograms, which represent the frequency distributions of the individual factors in a session. By using a large dataset, various insights into the relative impacts of these factors are then provided, serving as suggestions to improve the QoE of streaming sessions. 
	\end{abstract}

	\section{Introduction}\label{Sec_Intro}
	HTTP Adaptive Streaming (HAS) has become a popular solution for multimedia delivery nowadays. In HAS, a video is encoded into multiple versions with different bitrates (and so different quality values). Each version is further  divided into short segments~\cite{thang2012_TCE_full}. Based on network statuses, suitable versions of individual segments are selected and delivered to clients so that the highest possible quality of experience (QoE) can be provided to users. Towards effective version selections, a main challenge is to quantify the impacts of factors on the QoE of streaming sessions. 
	
\begin{figure}[!t]
		\centering
		\includegraphics[width=0.9\columnwidth]{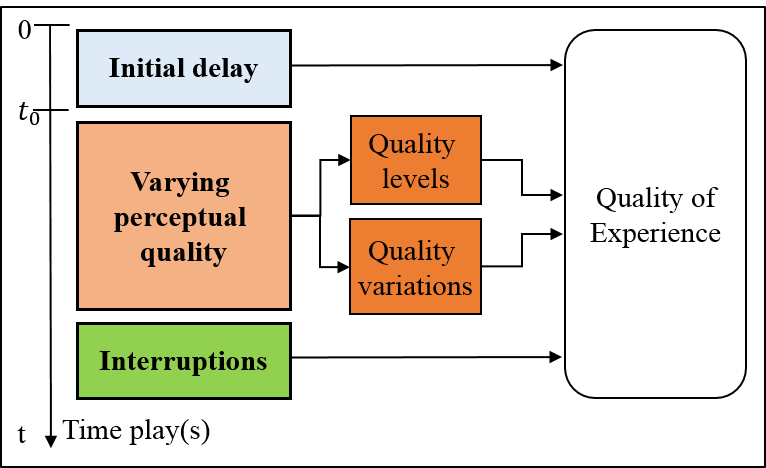}
		\caption{A taxonomy of key factors}
		\label{fig_tax}
	\end{figure}

	Previous studies have investigated, both qualitatively and quantitatively, different factors affecting the QoE of HAS sessions~\cite{QoE_xue2014_assessing,QoE_nam2016_qoewhycat,QoE_rodriguez2014_Brazil}. In general, there are three key factors, namely \textit{initial delay}, \textit{varying perceptual quality}, and \textit{interruptions} as shown in Fig.~\ref{fig_tax}. The initial delay refers to the waiting time before watching a video~\cite{QoE_tobias2012_initial}. Varying perceptual quality refers to quality changes of segments in a session as a consequence of network bandwidth fluctuations. This factor could be further divided into two sub-factors. The first, called \textit{quality levels}, refers to contributions of high and low segment quality levels on the QoE. The second, called \textit{quality variations}, refers to impacts of segment quality switches. Interruptions refer to instances of rebuffering while watching a video~\cite{QoE_tobias2012_initial}.

	In the literature, the impact of varying perceptual quality was modeled using some statistics such as the number of switches~\cite{QoE_bellLab2013_QoEmodel,QoE_shen2015_IEICE}, the average~\cite{QoE_bellLab2013_QoEmodel,QoE_rodriguez2014_Brazil,QoE_Yin_2015,QoE_bentaleb2016sdndash}, the median~\cite{QoE_ywang2015_assessing}, the minimum~\cite{QoE_ywang2015_assessing}, and the standard deviation of segment quality values~\cite{QoE_bellLab2013_QoEmodel}. As for interruptions, their impact was modeled using some statistics such as the number of interruptions~\cite{QoE_singh2012qualityML,QoE_liu2015_deriving}, the average~\cite{QoE_singh2012qualityML}, the maximum~\cite{QoE_singh2012qualityML}, and the sum~\cite{QoE_rodriguez2016_video,QoE_liu2015_deriving} of interruption durations. 
 	
 	Different from the two above factors, the impact of the initial delay was mostly found to be small~\cite{QoE_tobias2013_DTMA,QoE_seufert2015_survey,QoE_Survey2019}. In addition, as the initial delay appears only once at the beginning of a session, it is simple to individually model the impact of this factor using a function of the initial delay duration. This could be a linear function~\cite{QoE_liu2015_deriving}, an exponential function~\cite{QoE_rodriguez2016_video}, or a logarithmic function~\cite{QoE_tobias2012_initial,tran2016_GC}. Thus, in this study, we mainly focus on the two more important factors of varying perceptual quality and interruptions.

	Though previous studies~\cite{QoE_liu2015_deriving,tran2016_ICCE,QoE_tobias2012_initial,QoE_singh2012qualityML,QoE_rodriguez2016_video} have revealed some general behaviors of each individual factor, insights into relative impacts of different events (i.e., switching and interruptions) during a session are very limited. The understanding of such relative impacts can help make more effective adaptation decisions. For example, instead of abrupt down-switching to a very low quality value, remaining a moderate quality value with a very short interruption could bring a higher QoE. However, to the best of our knowledge, there are only two studies in \cite{QoE_egger2014_impact,QoE_nam2016_qoewhycat} that could give some findings on this issue.   In particular, the authors in~\cite{QoE_egger2014_impact} show that a down-switch can result in a comparable impact to an interruption. Meanwhile, an extensive study with YouTube users in~\cite{QoE_nam2016_qoewhycat} indicates that an interruption has three times the impact of a quality switch. Both the studies are not clear how different degrees of switches can be compared to different interruption durations, and so causing the confusion between the conclusions.  
	
	In this paper, our aim is to quantitatively investigate the relative impacts of different events of quality switching and interruption. In particular, we focus on answering four important questions below.
	\begin{itemize}
		\item How different are the impacts of different segment quality levels?
		\item What are the impacts of quality switching types? Do switches with higher switching amplitudes always cause more negative impacts? Does the starting quality of switches have any influence on the QoE? 
		\item How different are the effects of different interruption durations?
		\item What are the relative impacts between quality switching types and interruption types? Do interruptions always result in more negative effects than quality switches? 
	\end{itemize}

	For this purpose, we first extend our QoE model using histograms that has been presented in~\cite{tran2016_GC}. In particular, switches are classified into different types based on not only their \textit{switching amplitudes} (i.e., differences of segment quality before and after switching) but also \textit{starting quality values} (i.e., segment quality values before switching). Then, by using two more datasets, the reliability of the model is confirmed. Finally, based on an analysis of the model parameters, the impact of each switching or interruption type could be quantified. Also, various insights into their relative impacts are provided, which help answer the above questions.
	
	The remainder of the paper is organized as follows. Section~\ref{Sec_Related} presents the related work. The quality model, which aims at quantifying the relative impacts of different switching and interrupting events, is described in Section~\ref{Sec_model}. The performance of the proposed model is analyzed in Section~\ref{Sec:Performance}. Section~\ref{Sec_Dis} draws a set of insights into the impacts of the factors on the QoE. Finally, Section~\ref{Sec_Conc} concludes the paper.
	
	\section{Related Work}\label{Sec_Related}
	In this section, we first present related work on the impact of each single factor. Then, relative impacts of factors found in previous studies are described.  
	
	\subsection{Single Factors}\label{Sec_Related1}
	
	Aforementioned, the factor of varying perceptual quality could be divided into two sub-factors of quality levels and quality variations. The contributions of quality levels were investigated in many existing studies~\cite{QoE_bellLab2013_QoEmodel,QoE_rodriguez2014_Brazil,QoE_Yin_2015,QoE_bentaleb2016sdndash}. It was found that, given a quality level, its contribution depends on its total presence time during a session~\cite{QoE_tobias2014_assessing,QoE_liu2015_deriving}. To model these contributions, some statistics of segment quality values such as the average~\cite{QoE_bellLab2013_QoEmodel,QoE_rodriguez2014_Brazil,QoE_Yin_2015,QoE_bentaleb2016sdndash}, the median~\cite{QoE_ywang2015_assessing}, and the weighted sum ~\cite{QoE_liu2015_deriving} can be used.  
	
	There have been some findings on the impacts of quality variations presented in~\cite{QoE_nam2016_qoewhycat,QoE_tobias2014_assessing,QoE_tavakoli2016_JSAC,QoE_ni2011_flicker}. The experiment results in~\cite{QoE_nam2016_qoewhycat,QoE_ni2011_flicker} show that users expect the number of quality switches (or quality switching frequency) as low as possible. In contrast, the finding in~\cite{QoE_nam2016_qoewhycat} is that users prefer constant quality to varying quality. Even the impact of frequent quality switching may be more than four times higher than that with no quality change. Meanwhile, the impact of the number of quality switches is found to be negligible in~\cite{QoE_tobias2014_assessing,QoE_tavakoli2016_JSAC}. In particular, no significant difference was observed between high and low numbers of quality switches. The disagreement in the conclusions may stem from the fact that the use of the number of quality switches equitably treats all quality switch types while they may cause significantly different impacts. In other words, the conclusions are  specific to only the corresponding experiment results in the original papers, but not general in practice.    
	
	
	Some in-depth investigations on the impact of quality variations were conducted with classifying quality switches ~\cite{QoE_egger2014_impact,QoE_tobias2014_assessing,QoE_tavakoli2016_JSAC,QoE_liu2015_deriving}. It is found that the impacts of up-switches are much smaller than those of down-switches~\cite{QoE_liu2015_deriving}. In addition, abrupt up-switches may not be worse than smooth up-switches~\cite{QoE_egger2014_impact,QoE_tavakoli2016_JSAC}. Meanwhile, down-switches with larger switching amplitudes cause more negative impacts~\cite{QoE_tobias2014_assessing,QoE_tavakoli2016_JSAC}. Similar conclusions are also given in~\cite{tran2017_IEICEhistogram}, where the authors quantify the impacts of switching types with different switching amplitudes. However, these findings are limited because it cannot differentiate switches having different starting quality values (e.g., a switch from 5~MOS to 3~MOS is in fact not the same as a switch from 3~MOS to 1~MOS). So far, there has been no study on the impact of starting quality values of switches. 
	
	In most existing studies, the impact of quality variations are modeled using some statistics such as the  number of switches~\cite{QoE_bellLab2013_QoEmodel,QoE_shen2015_IEICE}, the minimum~\cite{QoE_ywang2015_assessing}, and the standard deviation of segment quality values~\cite{QoE_bellLab2013_QoEmodel}. In our previous study~\cite{tran2016_GC}, it is found that the use of histograms of switching amplitudes is very effective to model the impact of this sub-factor. 
	
	With regard to interruptions, the authors in~\cite{QoE_nam2016_qoewhycat} showed that users prefer a single interruption to multiple interruptions. The impact of a single interruption is modeled as an exponential function of its duration in~\cite{QoE_tobias2012_initial}. To model the impacts of multiple interruptions, several previous studies used the number of interruptions~\cite{QoE_singh2012qualityML,QoE_liu2015_deriving}, the average~\cite{QoE_singh2012qualityML}, the maximum~\cite{QoE_singh2012qualityML}, and the sum~\cite{QoE_rodriguez2016_video,QoE_liu2015_deriving} of interruption durations.

	\subsection{Relative Impacts of Factors}
	
	Recently, many QoE models have been proposed for HAS. However only a few are multi-factor models~\cite{QoE_liu2015_deriving,QoE_rodriguez2016_video}. The studies in~\cite{QoE_QoEIndex_ZDuanmu2018,QoE_XLiu2012,QoE_xue2014_assessing,QoE_singh2012qualityML} modeled the contributions of quality levels and interruptions. However, the impact of quality variations was not considered. In~\cite{QoE_rodriguez2016_video}, the authors proposed a model taking into account the impacts of quality variations and interruptions. Yet, this model does not include the impacts of quality levels. 
	
	To the best of our knowledge, the authors in~\cite{QoE_liu2015_deriving} proposed the first QoE model taking into account all the three (sub-)factors of quality levels, quality variations, and interruptions. This model is built in two steps, which are 1) separately modeling and then 2) combining the impacts of factors. In particular, the impact of a quality level is modeled by the total presence time of that quality level. For the impacts of quality variations, the authors used the mean square of down-switching amplitudes. The impact of interruptions is modeled using the number of interruptions and the sum of interruption durations. 	In the latest stage of standardization \itut~\itutp~\cite{ITU1203_3}, a multi-factor model is recommended. In this model, the impacts of quality levels,  quality variations, and interruptions are modeled using various statistics such as the difference between the maximum and minimum segment quality values, the weighted sum of segment quality values, and the number of interruptions. 
	
	
	
	Although the factors were modeled in the above studies, the relative impacts of these factors have not been quantified. In the literature, there are very few studies investigating this issue. In~\cite{QoE_egger2014_impact}, the finding is that a quality switch can cause a comparable impact as an interruption. However, the switching and interruption types considered in that study are limited. In particular, only two specific pairs consisting of two switching types and two interruption types were compared in that study. An study in~\cite{QoE_nam2016_qoewhycat} revealed that an interruption has three times the impact of a switch. However, the switching and interruption types are not clearly defined. 
	
	In this study, we, for the first time, attempt to fully quantify the impacts of different switching and interruption types. Based on obtained results, a set of insights into the relative impacts of the factors are provided. 
		
	\section{Proposed QoE model}\label{Sec_model}
	In this section, we first present two main components of the proposed model. The first, denoted $Q_{PQ}$, represents the impact of varying perceptual quality. The second, denoted $D_{IR}$, represents the impact of interruptions. Then, a combination of these components to predict the QoE is given.
	
	\subsection{Impact of Varying Perceptual Quality}
	To model the impact of varying perceptual quality, we utilize two histograms of two sub-factors, i.e., quality levels and quality variations.
	
	In particular, each segment is represented by a quality value (i.e., MOS), which can be obtained by subjective tests~\cite{QoE_bellLab2013_QoEmodel,QoE_tobias2014_assessing,tran2016_GC} or estimated from encoding parameters~\cite{QoE_ywang2015_assessing,tran2016_ICCE}. The range of segment quality values is split into $N$ intervals ${\{I_n^Q|1 \le n \le N\}}$, which are given by  
	\begin{equation}
	I_n^Q=[n-\vartheta_{L_n},n+\vartheta_{U_n } ),
	\end{equation}
	where $\vartheta_{U_n}$  and $\vartheta_{L_n}$ are parameters to define the intervals' widths.
	Each interval $I_n^Q$ corresponds to a segment quality bin $B_n^Q$. If a segment quality value is in interval $I_n^Q$, it belongs to bin $B_n^Q$.   
	
	In the proposed model, each quality switch is represented by a starting quality value and a switching amplitude. To define switching amplitudes, we use the concept of \textit{``quality gradient''}, which is given by 
	\begin{equation}
	\nabla Q=  \partial Q/\partial t, 	
	\end{equation}
	where $\partial Q$ is the change of segment quality values in time interval $\partial t$. 
	
	Currently, we use the quality value of the segment just before switching to represent the starting quality value, and quality changes between two segments right before and after
	switching to represent the instant gradient value at each switch. A positive (negative) gradient value indicates an up-switch (down-switch). The range of gradient values is split into $(2\times M+1)$ intervals ${\{I_j^V|-M \le j \le M\}}$, which are defined by 
	\begin{equation}
	I_j^V=[j-\theta_{L_j},j+\theta_{U_j } ),
	\end{equation}
	where the parameters $\theta_{U_j }$ and $\theta_{L_j}$ define the intervals' widths. 
	
	Each quality switching bin $\{B_{i,j}^V (1 \le i \le N,$ $-M \le j \le M)\}$ is defined by two intervals $I_i^Q$ and $I_j^V$. A switch belongs to bin $B_{i,j}^V$ if its starting quality value is in interval $I_i^Q$ and its switching amplitude is in interval $I_j^V$. 
		
	Let $F_n^Q$ denote the normalized frequency of segment quality values in bin $B_n^Q (1 \le n \le N)$. $F_{i,j}^V$ denotes the normalized frequency of switches in bin ${B_{i,j}^V  (1 \le i \le N, -M \le j \le M)}$. Note that $F_n^Q$ is normalized by the number of segments. $F_{i,j}^V$ is normalized by the total number of switches and interruptions. 
	The perceptual quality $Q_{PQ}$ is modeled by
	\begin{equation}
	Q_{PQ}=\sum_{n=1}^{N}\alpha_n F_n^Q - \sum_{i=1}^N \sum_{j=-M}^M \beta_{i,j} F_{i,j}^V, 	
	\end{equation}
	where $\alpha_n$  and $\beta_{i,j}$ are respectively the weights of bin $B_n^Q$ and  bin $B_{i,j}^V$. 
	
	In this study, we use the Absolute Category Rating method with a 5-grade scale, which is widely used for quality assessments of streaming sessions in HAS~\cite{ITU1203_3,QoE_egger2014_impact,QoE_tavakoli2016_JSAC,QoE_robitza2017_modular}. So we currently split the range of segment quality values into $N=5$ bins, and the range of gradient values into 9 intervals $(M=4)$ with $\vartheta_{U_n }=\vartheta_{L_n}=\theta_{L_j}=\theta_{U_j} =0.5$.

	It can be seen that  intervals $\{I_j^V|j>0\}$ contain up-switches,  interval $I_0^V$ represents quality maintaining, and intervals $\{I_j^V|j<0\}$ include down-switches. 
	As noted in the previous study of~\cite{tran2017_IEICEhistogram}, the impacts of down-switches are significant while the impacts of non-negative switches (including up-switches and quality maintaining) are negligible. So, we simplify the proposed model by grouping all the bins of non-negative switches into one bin (denoted by $B^{um}$). The normalized frequency $F^{um}$ of this bin is given by 
	\begin{equation}
	F^{um}=\sum_{i=1}^N \sum_{j=0}^M F_{i,j}^V.
	\end{equation}
	Then, the simplified perceptual quality model is given by 
	\begin{equation}
	Q_{PQ}=\sum_{n=1}^N \alpha_n F_n^Q - \sum_{i=1}^N \sum_{j=-M}^{-1} \beta_{i,j} F_{i,j}^V - \beta^{um} F^{um},
	\end{equation}
	where $\beta^{um}$  is the weight of bin $B^{um}$. With the simplified model, the number of model parameters (i.e., weights)  can be reduced by approximately a half.

	\subsection{Impact of Interruptions}

	To investigate the impact of interruptions, a histogram of this factor is defined as follows. Each interruption is represented by its duration. The range of interruption durations is divided into $L$  intervals $\{I_l^I|1 \le l \le L\}$ corresponding to $L$ interruption bins $\{B_l^I|1 \le l \le L\}$. 
	Let $F_l^I$ be the normalized frequency of interruptions in bin $B_l^I$. 
	Note that $F_l^I$ is  normalized by the total number of switches and interruptions. 
	The impact of interruptions $D_{IR}$ is modeled by 
	\begin{equation}
	D_{IR}= \sum_{l=1}^L \gamma_l F_l^I,
	\end{equation} 
	where $\gamma_l$ is the weight of bin $B_l^I$.

	\begin{figure}[!t]
		\centering
		\subfloat[]{\includegraphics[width=0.48\columnwidth]{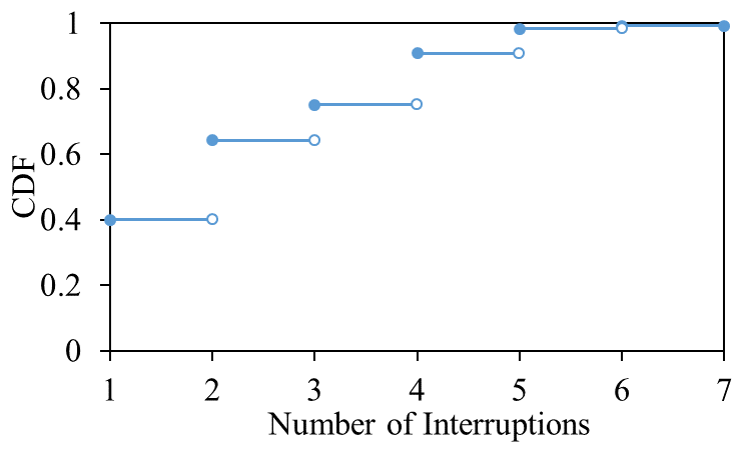}%
			\label{fig_Dis_Intera}}
		\hfil
		\subfloat[]{\includegraphics[width=0.48\columnwidth]{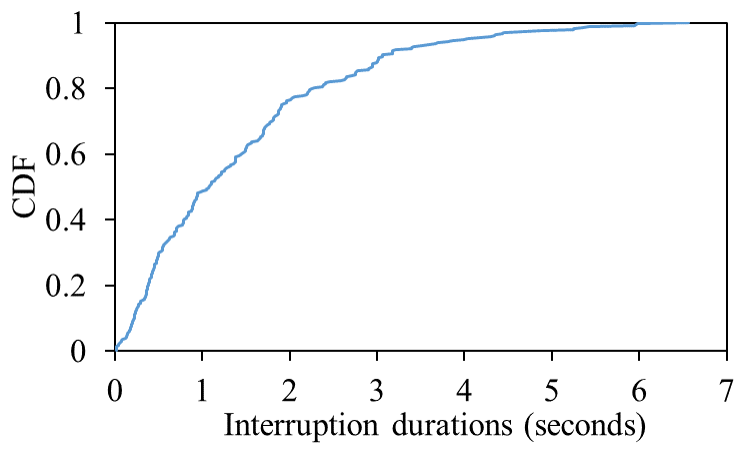}%
			\label{fig_Dis_Interb}}
		\caption{Distribution of interruptions: (a) left, the numbers of interruptions per session; (b) right, interruption durations.}
		\label{fig_Dis_Inter}
	\end{figure}
	
	\begin{table}[!t]
		\caption{Intervals of Interruption Durations.}
		\label{table_interval_inter}
		\resizebox{\columnwidth}{!}{%
		\centering
		\begin{tabular}{|c|c|c|c|c|c|}
			\hline 
			\multicolumn{6}{|c|}{\rule{0pt}{8pt}  \textbf{Interval (s)}}\\ [0.05cm]   \hline &&&&& \\[-1em]
			\textbf{$I_1^I$ } &	$I_2^I$ & $I_3^I$ &	$I_4^I$ &	$I_5^I$ &	$I_6^I$\\ \hline &&&&& \\[-0.7em]
			$(0.0,0.25]$ &	$(0.25,0.5]$ & $(0.5,1.0]$ & $(1.0,2.0]$ & $(2.0,3.0]$& $(3.0,+\infty)$\\
			\hline
		\end{tabular}}
	\end{table}

	\begin{table}[!t]
		\caption{Features of Source Videos}
		\label{table_features}
		\centering
		\resizebox{0.45\textwidth}{!}{%
			\begin{tabular}{|c|M{0.55in}|c|m{0.8in}|M{0.45in}|M{0.65in}|}
				\hline
				\textbf{Video} & \textbf{Frame rate (fps)} &\textbf{Type} & \centering \textbf{Content} & \centering \textbf{Motion activity} & \multicolumn{1}{M{0.65in}|}{\centering \textbf{Spatial complexity}}\\\hline
				\centering \textbf{\textit{Video \#1}}  &24& Animation & Slow movements of characters & Low & Complex\\\hline
				\centering \textbf{\textit{Video \#2}}  &24& Animation &A fight between two characters & High & Simple\\\hline
				\centering \textbf{\textit{Video \#3}} &24& News & A reporter in a weather forecast & Medium &Complex\\\hline 
				\centering \textbf{\textit{Video \#4}} &24& Sport &A soccer match &High &Complex\\\hline
			\end{tabular}
		}
	\end{table}
	
	To define the representative intervals $I_l^I$, a series of video streaming experiments was performed on a real testbed. In these experiments, we recorded totally 120 streaming sessions with at least one interruption per session. The distributions of the number of interruptions and interruption durations are shown in Fig.~\ref{fig_Dis_Inter}. We can see that, the number of interruptions per session is typically from 1 to 6. About 40\% of the sessions contains only one interruption. Besides, about 13\% of the observed interruptions has durations shorter than 0.25 seconds, and about 10\% longer than 3 seconds. Based on these observations, we split the range of interruption durations into $L = 6$ intervals as shown in Table~\ref{table_interval_inter}. Note that the special interval $I_6^I$ includes interruption durations longer than 3 seconds.
	

	\subsection{Overall QoE Model}
	
	Finally, the proposed model, which integrates the above components to predict the QoE of streaming sessions, is given by 
	\begin{equation}
	QoE_{pred}= \max(Q_{PQ}-D_{IR},1).
	\end{equation}
	This model has totally 22 model parameters which enable comparisons of impacts between switching and interruption types presented in Section~\ref{Sec_Dis}.

	\section{Performance Evaluation}\label{Sec:Performance}
	
	To evaluate the prediction performance of the proposed model, we combine our two databases presented in~\cite{tran2016_GC,tran2019_INFOCOM}. The combination database consists of 288 sessions, of which 168 contain only one single factor (i.e., either varying perceptual quality or interruption) and 120 include both the factors. From this database, we choose 50 pairs of training and test sets. In particular, for each pair, the test set consists of 90 sessions randomly selected from the multi-factor sessions. The corresponding training set consists of 198 remaining sessions (i.e., 168 single-factor sessions and 30 remaining  multi-factor sessions). The training set is to obtain model parameters using least squares fitting. The test set is to evaluate prediction performances. The results presented in the following are the average values over the 50 test sets.

	\begin{table}[!t]
		\caption{Performance of the Proposed QoE Model}
		\label{table_per}
		\centering
		\begin{tabular}{|m{0.62in}|M{0.25in}|M{0.35in}|M{0.25in}|M{0.35in}|}
			\hline
			\multirow{2}{*}{\centering \textbf{Model}} & \multicolumn{2}{c|}{\textbf{Training set}} & \multicolumn{2}{c|}{\textbf{Test set}}\\ \cline{2-5}
			& PCC & RMSE & PCC & RMSE \\ \hline
			\textbf{\textit{Proposed}} & 0.97 & 0.27 & 0.95 & 0.30 \\ \hline
		\end{tabular}
	\end{table}

	Table~\ref{table_per} shows the performance of the proposed model in terms of Person Correlation Coefficient (PCC) and Root Mean Squared Error (RMSE). It can be seen that the proposed model achieves very high PCC values and very low RMSE values. In particular, the PCC and the RMSE values are (0.97, 0.27) for the training sets and (0.95, 0.30) for the test sets. This indicates that the proposed model can predict well the QoE of streaming sessions with the impacts of varying perceptual quality and interruptions. 
	
	\subsection{Model Parameters}\label{Sec:Par}
	Similar to~\cite{QoE_liu2015_deriving}, Tables~\ref{table_para}, \ref{table_para_sw}, and \ref{table_para_inter} show the values of the model parameters $\{\alpha_n,\beta_{i,j},\gamma_l\}$ corresponding to the test set which achieves the highest PCC value (i.e., 0.96) among the 50 test sets. It is interesting to see that the weights $\alpha_n$ and $\gamma_n$ increase w.r.t. their indexes of $n$ and $l$, respectively. The weight $\beta^{um}$ of the non-negative switching bin $B^{um}$ is equal to zero. For down-switching bins $\{B^V_{i,j}|j<0\}$, their weights $\{\beta_{i,j}\}$ increase when the switching amplitudes get larger values or the starting quality values become lower. More detailed discussions about these weights will be made in Section~\ref{Sec_Dis}.

	\begin{table*}[!t]
		\centering
		\caption{Description of the proposed and existing models}
		\label{Table:DescRefModel}
		\resizebox{2\columnwidth}{!}{%
		\begin{tabular}{|m{0.65in}|m{3.4in}|m{2.7in}|}
			\hline
			\centering \multirow{3}{*}{\textbf{Models}}  & \multicolumn{2}{c|}{\multirow{2}{*}{\textbf{Statistics used to represent the impacts of factors}}} \\ 
			&\multicolumn{2}{c|}{}  \\\cline{2-3}
			& \centering \addstackgap{\textit{\textbf{Varying Perceptual Quality}}}   &  \multicolumn{1}{c|}{\textit{\textbf{Interruptions}}}  \\  \hline
			\textbf{\textit{Guo's}}~\cite{QoE_ywang2015_assessing}   & \addstackgap{\shortstack[l]{Median segment quality value \\Minimum segment quality value}} &  \multicolumn{1}{c|}{\textemdash}  \\ \hline
			\textbf{\textit{Vriendt's}}~\cite{QoE_bellLab2013_QoEmodel}	& \addstackgap{\shortstack[l]{Number of switches \\ Average and standard deviation of segment quality values}}  &  \multicolumn{1}{c|}{\textemdash}\\ \hline
			\textbf{\textit{Liu's}}~\cite{QoE_liu2015_deriving}	 & \addstackgap{\shortstack[l]{Weighted sum of segment quality values\\Average of the squares of down-switching amplitudes}}  &  \shortstack[l]{Sum of interruption durations \\ Number of interruptions}\\ \hline
			\textbf{\textit{\itutp}}~\cite{ITU1203_3}  & \addstackgap{\shortstack[l]{Number of switches \\ Number of quality direction changes \\ Longest switching duration\\ First and fifth percentile of segment quality values \\ Average of segment quality values in each interval \\ Weighted sum of segment quality values \\Difference between the maximum and minimum of segment\\ quality values}} &  \addstackgap{\shortstack[l]{Number of interruptions\\ Weighted sum of interruption durations\\ Average time distance between interruptions \\ Sum of interruption durations \\ Time distance between the last interruption and\\ the end of session }}  \\ \hline
			\textbf{\textit{Proposed}}& \addstackgap{\shortstack[l]{Histogram of segment quality values \\ Histogram of switching amplitudes}} & Histogram of interruptions   \\ \hline
		\end{tabular}}
	\end{table*}

	\begin{table*}[!t]
		\centering
		\caption{Adjustment Coefficients and Performances of the Proposed Models and Existing Models}
		\label{table_perAll}
		\resizebox{1.7\columnwidth}{!}{%
		\begin{tabular}{|m{0.55in}|M{0.40in}|M{0.50in}|M{0.40in}|M{0.40in}|M{0.40in}|M{0.50in}|M{0.40in}|M{0.40in}|}
			\hline
			\multirow{3}{*}{\centering \textbf{Model}}  & \multicolumn{4}{c|}{\textbf{Our database}} & \multicolumn{4}{c|}{\textbf{VL04}}\\ \cline{2-9}
			& \multicolumn{2}{c|}{\textit{\textbf{Coefficients}}}  & \multicolumn{2}{c|}{\textit{\textbf{Performance}}} & \multicolumn{2}{c|}{\textit{\textbf{Coefficients}}}  & \multicolumn{2}{c|}{\textit{\textbf{Performance}}}  \\ \cline{2-9}
			& \textit{Slope}  & \textit{Intercept}  & \textit{PCC} & \textit{RMSE}& \textit{Slope}  & \textit{Intercept}  & \textit{PCC} & \textit{RMSE}  \\ \hline
			\textit{\textbf{Guo's}} &0.53& 0.84&  0.69 & 0.61& 0.75 & 0.67 & 0.72 & 0.62\\ \hline
			\textit{\textbf{Vriendt's}}&0.56&0.71&  0.77 & 0.58& 0.81& 0.50  & 0.71& 0.63 \\ \hline
			\textit{\textbf{Liu's}}&1.18&-1.32& 0.78 & 0.56& \textemdash & \textemdash & \textemdash& \textemdash \\ \hline
			\textit{\textbf{P.1203}} &1.14 &-1.30&  0.85 & 0.42& 1.04 & 0.08 & 0.88& 0.42 \\ \hline
			\textit{\textbf{Proposed}} &\textemdash&\textemdash&0.95 & 0.30& 0.79  & 0.82 & 0.90 & 0.39 \\ \hline
		\end{tabular}}
	\end{table*}

	\subsection{Model Comparisons }\label{Sec:Compare}

	In this part, a comparison of prediction performance between the proposed model and four existing models is conducted over two databases. The first is our database, where prediction performances are the average values over the 50 test sets mentioned in Subsection~\ref{Sec:Performance}. The second is an open database (called \textit{VL04}) of the \itut~\itutp standardization procedure (P.NATS)~\cite{ITUT_dataset,ITUT_implement2}. This database consists of sixty 1-minute long sessions generated from three different videos. The performances of the models are calculated over all these sixty sessions. 
	
	A description of the proposed model and the four existing models, denoted by \textit{Guo's}~\cite{QoE_ywang2015_assessing}, \textit{Vriendt's}~\cite{QoE_bellLab2013_QoEmodel}, \textit{Liu's}~\cite{QoE_liu2015_deriving}, and \textit{\itutp}~\cite{ITU1203_3,ITUT_implement1,ITUT_implement2,ITUT_implement3}, is presented in Table~\ref{Table:DescRefModel}. It can be seen that these models use various statistics to model the impacts of the factors. The models \textit{Guo's} and \textit{Vriendt's} only take into account the impacts of varying perceptual quality. Meanwhile, the remaining models include the impacts of both varying perceptual quality and interruptions. 
	
	According to Recommendations \itut~P.1401~\cite{ITUT_Rec1401} and \itut~\itutp~\cite{ITU1203_3}, we conducted a compensation for test condition differences between models. 
	In particular, each reference model is re-implemented by using parameters stated in the original study. Then, a first-order linear regression is performed to adjust the predicted QoE values. Finally, after the adjustment, the prediction performance is calculated. The coefficients (i.e., slopes and intercepts) of the regression and performance corresponding to each database are presented in Table~\ref{table_perAll}.

	It can be seen that, with using our database, the proposed model achieves the best prediction performance (i.e., the highest PCC value and the lowest  RMSE value). Specifically, the PCC and RMSE values of the proposed model are respectively 0.95 and 0.30~MOS. It is clear that the models \textit{Guo's} and \textit{Vriendt's} fail to predict the QoE values of multi-factor sessions since they do not include the impacts of interruptions. Thus, their PCC values are low (i.e., $\le 0.77$), and their RMSE values are very high (i.e., $\ge 0.58$~MOS).  For the models \textit{Liu's} and \textit{\itutp}, their performances are significantly lower than that of the proposed model. In particular, the PCC and RMSE values are respectively 0.78 and 0.56~MOS for the model \textit{Liu's}, and 0.85 and 0.42~MOS for the model \textit{\itutp}. A possible explanation for this result is that these models use some statistics such as the number of switches and the number of interruptions to model the impacts of varying perceptual quality and interruptions. However, these statistics can not fully reflect switches and interruptions occurring in a session as they can not distinguish different switching types and interruption types. Meanwhile, thanks to the use of the histograms, the proposed model can differentiate different switching and interruption types, and so can more effectively model the impacts of the factors.

	\section{Analysis of relative impacts}\label{Sec_Dis}

	\begin{table}[!t]
		\caption{Parameters of the Segment Quality Component}
		\label{table_para}
		\centering
		\begin{tabular}{|M{0.32in}|M{0.32in}|M{0.32in}|M{0.32in}|M{0.32in}|}
			\hline
			$\alpha_1 $ & $\alpha_2$ & $\alpha_3$ & $\alpha_4$ & $\alpha_5$\\\hline
			1.11  & 2.20  & 3.20 & 4.00 & 4.50\\\hline
		\end{tabular}
	\end{table}	
	
		\begin{table}[!t]
		\caption{Parameters of the Quality Switching Component}
		\label{table_para_sw}
		\centering
		\resizebox{\columnwidth}{!}{%
		\begin{tabular}{|M{0.7in}|M{0.55in}|M{0.27in}|M{0.27in}|M{0.27in}|M{0.27in}|}
			\hline
			\multicolumn{2}{|c|}{\centering \multirow{2}{*}{\centering \textbf{$\beta_{i,j}$}}} &  \multicolumn{4}{c|}{\centering \textbf{Starting quality value (i)}} \\ \cline{3-6}
			\multicolumn{2}{|l|}{}&5& 4&3&2 \\\hline
			\multirow{5}{*}{\addstackgap{\shortstack[l]{\textbf{{~~~~Quality}} \\\textbf{{~~switching}}\\ \textbf{{amplitude}}\\\textbf{{~~~~~~~(j)}} }}}  & \centering	\textit{Non-neg. ($\beta^{um}$)}& \multicolumn{4}{c|}{\centering 0.0}\\ \cline{2-6}
			 &\centering	\textit{-1}&	0.01	&0.01	&3.93	&7.89 \\ \cline{2-6}
			 &\centering	\textit{-2}&	3.93	&4.13	&14.36&	\textemdash \\ \cline{2-6}
			 &\centering	\textit{-3}&	18.69&	18.99& \multicolumn{2}{c|}{\centering \textemdash} \\ \cline{2-6}
			&\centering	\textit{-4}&	24.76&	\multicolumn{3}{c|}{\centering \textemdash} \\ 	\hline
		\end{tabular}}
	\end{table}
	\begin{table}[!t]
		\caption{Parameters of the Interruption Component}
		\label{table_para_inter}
		\centering
		\begin{tabular}{|M{0.32in}|M{0.37in}|M{0.32in}|M{0.32in}|M{0.32in}|M{0.32in}|}
			\hline
			$\gamma_1$  & $\gamma_2$ &$\gamma_3$  & $\gamma_4$ & $\gamma_5$& $\gamma_6$  \\\hline
			0.00	&8.42	&16.15	&24.16	&45.58	&50.65 \\\hline
		\end{tabular}
	\end{table}
	
	\subsection{Impacts of Factors}
 	In this subsection, we quantitatively analyze the impacts of the factors by discussing the weights of the bins in the proposed model.  
		
	Firstly, we give analysis of the weights $\alpha_n$ of the segment quality bins $B^Q_n$ shown in Table~\ref{table_para}. It can be seen that these weights increase with their indexes $n$. This implies that a higher quality level has bigger contribution in the QoE of sessions. In other words, the contribution of the highest quality level is biggest, which is similar to the finding in~\cite{QoE_tobias2014_assessing}. Note that, only two quality levels are considered in~\cite{QoE_tobias2014_assessing}. It is interesting to see that most of these weights are similar to the midpoints of the corresponding intervals $I^Q_n$, except for the weight $\alpha_5$. The reason may be because it is, in fact, difficult to achieve 5~MOS even at perfect quality~\cite{QoE_tominaga2010_performance,QoE_tobias2012_initial}. 
	
	Next, we quantitatively investigate the impacts of quality variations. Table~\ref{table_para_sw} reports the weights $\beta_{i,j}$ of the quality switching bins $B^V_{i,j}$. It is found that the impacts of switches depend not only on switching amplitudes but also on  starting quality values. With the same starting quality value, the larger the switching amplitude is, the more serious the impact becomes. Also, given a switching amplitude, the lower the starting quality value is, the more negative the impact is. This means, when having the same switching amplitude, a down-switch in low quality ranges is more severe than in high quality ranges. For example, as $\beta_{3,-2} > \beta_{4,-2}$ and $\beta_{4,-2} > \beta_{5,-2}$, a switch from 3~MOS to 1~MOS has more negative impact than that from 4~MOS to 2~MOS, which is in turn worse than from 5~MOS to 3~MOS. 
	
	Although the switching amplitude is smaller, $\beta_{2,-1}$ is higher than $\beta_{5,-2}$ and $\beta_{4,-2}$. This reveals that switches having larger switching amplitudes may not cause more negative impacts. From Table~\ref{table_para_sw}, it is also observed that the weights of $\beta_{2,-1},\beta_{3,-2},\beta_{4,-3},$ and $\beta_{5,-4}$ (in increasing order) are very large (up to 24.76). So it is recommended to avoid switching to very low quality levels (i.e., around 1$\sim$1.5~MOS), especially from a high starting quality value.

	
	In addition, the weight $\beta^{um}$ turns out to be zero. This reconfirms the finding in~\cite{tran2017_IEICEhistogram} that the impacts of up-switches and quality remaining are negligible. Therefore, it is unnecessary to classify up-switches (like down-switches). This finding is in line with those obtained in~\cite{QoE_egger2014_impact,QoE_tavakoli2016_JSAC} that abrupt up-switches may not be worse than smooth up-switches. Even users prefer switching up to and then maintaining a high quality level to gradual switching.  
	In addition, this result implies that switching to higher quality levels is preferred to remaining at low quality levels, which is in agreement with~\cite{QoE_tavakoli2016_JSAC}. 
	
	In contrast, a study in~\cite{QoE_nam2016_qoewhycat} shows that frequent quality increasing can lead to significantly larger impacts compared to no quality change. This observation can be obtained when comparing a session having frequent up-switches in low quality ranges with a session having quality fixed at higher levels. However, this conclusion may not be correct for up-switching in high quality ranges and quality remaining at low levels. Therefore, to avoid confusions, findings should be specific to switching types.	
	
	\begin{figure}[!t]
		\centering
		\includegraphics[width=0.8\columnwidth]{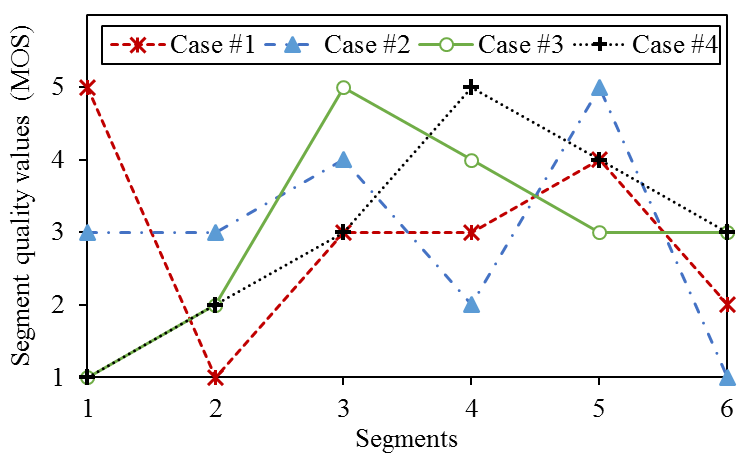}
		\caption{An example of four sessions with different quality variations.}
		\label{fig_QV_Disc}
	\end{figure}
	
	To better understand some statistics of segment quality values, Fig.~\ref{fig_QV_Disc} shows some cases of sessions with different quality variations. All the cases have the same statistics including the average, the median, the maximum, the minimum, and the standard deviation of segment quality values. However, it can see that quality variations are very different in these cases. So these statistics are not able to fully reflect quality variations occurring in a session. Although the number of switches in the case \#4 is higher than in the cases \#1, \#2, and \#3, the predicted QoE value (from the proposed model) of the case \#4 is similar to that of the case \#3, but significantly higher than that of the cases \#1 and \#2. It is because that, in both of the cases \#3 and \#4, down-switches have small amplitudes, and there is no down-switch to very low quality levels. Meanwhile, the cases \#1 and \#2 have down-switches from 5~MOS to 1~MOS, which cause very serious impacts. This result suggests that the larger number of switches may not lead to the lower QoE value.

	Then, we discuss the weights $\gamma_l$ of the interruption bins $B^I_l$ shown in Table~\ref{table_para_inter}. We can see that, the larger the index $l$ is, the higher the weight $\gamma_l$ becomes. This means that the impact of interruptions increases with their durations. Especially, the weight $\gamma_0$ is zero, implying that users are generally tolerant of interruptions having durations less than or equal to 0.25~seconds. For interruptions with longer durations, their impacts are much more negative.

	In a comparison between the weights $\beta_{i,i}$ and $\gamma_l$, it can also be observed that the weights $\gamma_5$ and $\gamma_6$ are very large (45.58 and 50.65), even larger than the weights $\beta_{i,j}$ of any quality switching bins. This indicates that an interruption with duration longer than 2 seconds is more annoying to users than any down-switches. Therefore, avoiding such interruptions should be of the highest priority, possibly at the cost of abrupt switches. 
	
	Meanwhile, the weight $\gamma_2$ is close to the weight $\beta_{2,-1}$ (i.e., 8.42 and 7.89). This result is in-line with~\cite{QoE_egger2014_impact} that a down-switch may cause a comparable impact as an interruption. Moreover, it can be seen that the weight $\gamma_3$ is considerably lower than the weights $\beta_{5,-3},\beta_{4,-3},$ and $\beta_{5,-4}$. This indicates that a down-switch can even result in more serious impacts than an interruption. In other words, an interruption does not necessarily result in more negative impacts than a down-switch. 
	
	In~\cite{QoE_nam2016_qoewhycat}, the finding is that an interruption has three times the impact of a switch. However, the types of interruptions and switches are not clearly mentioned. It can be seen that the weight $\gamma_4$ (i.e., 24.16) is approximately three time of the weight $\beta_{2,-1}$ (i.e., 7.89). However, for different weight pairs, the ratio considerably changes. This again shows that findings should be specific to switching and interruption types.	
	 	   
	Similar to switching types, different interruption types have different weights. Therefore, the number of interruptions, the average, the maximum, and the sum of interruption durations are also not able to fully reflect interruptions occurring in a session.  
	 
	Finally, it can be seen that the above findings are enabled by the use of histograms in the proposed model. In this way, it is possible to provide a set of insights into the relative impacts of the different factors. In other words, the use of the histograms is more flexible and comprehensive than the use of some statistics such as the average, the median, the minimum, and the standard deviation. 
	
	\subsection{Remarks of Results}
	Based on the above results and discussions, some remarks on the impacts of the factors can be summarized as follows.
	\begin{itemize}
		\item  First, it is found that a higher quality level has bigger contribution in the QoE of sessions.
		\item  Second, the effects of switches depend on not only switching amplitudes but also starting quality values. So switches having larger amplitudes do not necessarily cause more negative impacts.
		\item  Third, it is suggested that switching to a very low quality value (i.e., around 1$\sim$1.5~MOS) should be avoided, especially from a high starting quality value.
		\item  Fourth, the impact of interruptions increases with their duration. Interruptions having durations less than or equal to 0.25 seconds have trivial influences. Meanwhile, interruptions having durations longer than 2 seconds should be avoided with the highest priority, possibly at the cost of abrupt switches.  
		\item 	Fifth, an interruption does not necessarily result in more negative impacts than a down-switch.   
		\item 	Finally, the impacts of varying perceptual quality and interruptions can be effectively modeled by using histograms. 
	\end{itemize}

	\section{Conclusions}\label{Sec_Conc}
	In this paper, we have first proposed a QoE model taking into account the impacts of varying perceptual quality and interruptions. Then, by using the two databases, the experiment results have showed that the proposed model has high prediction performance and outperforms the four existing models. Finally, by discussing the model parameters, a set of the insights into the impacts of the factors have been provided in detail to each switching and interruption type. We hope that the findings in this paper will be useful for researchers in better understanding the factors affecting the QoE of streaming sessions, and further providing some suggestions to improve adaptation strategies in HTTP Adaptive Streaming.   
	
	For future work, we will extend the propose model to take into account the impact of the initial delay. Besides, we will seek to apply the proposed model in evaluations and developments of adaptation strategies, such that they can utilize network resources to provide the best quality of experience.

	
	
	\bibliographystyle{IEEEtran}
	\bibliography{Citation_Stand_full}
	%

\end{document}